\documentclass[12pt]{article}

\usepackage{a4wide}

\usepackage{amsmath}
\usepackage{amscd}
\usepackage{amsfonts}
\usepackage{amssymb}
\usepackage{mathrsfs}
\usepackage{fancybox} 
\usepackage{enumerate}
\usepackage{multirow}

\usepackage{cite}
\usepackage{bm}
\usepackage{amscd}
\usepackage{graphicx}
\usepackage[dvips]{color}
\usepackage{epsfig}
\usepackage{subfigure}

\makeatletter

\@addtoreset{equation}{section}
\makeatother
\usepackage{amsthm} 

\def\half{{1\over2}}

\def\const{\mathrm{const}}

\def\={\stackrel{\bullet}{=}}

\def\({\left(}
\def\){\right)}
\def\[{\left[}
\def\]{\right]}

\def\cO{{\cal O}}

\def \be {\begin{equation}}
\def \ee {\end{equation}}
\def \beqa {\begin{eqnarray}}
\def \eeqa {\end{eqnarray}}
\def \beal#1 {\begin{align}#1\end{align}}
\def \bes#1 {\begin{equation}\begin{split}#1\end{split}\end{equation}}
\def \nn {\notag\\}

\makeatletter

\@addtoreset{equation}{section}
\makeatother

\begin{document}

\begin{titlepage}
\title{
\vspace{-2cm}
\vspace{1.5cm}
Relativistic Hydrostatic Structure Equations\\
and Analytic Multilayer Stellar Model 
\vspace{1.cm}
}
\author{
Shuichi Yokoyama\thanks{syr18046[at]fc.ritsumei.ac.jp},\; 
\\[25pt] 
${}^{*}$ {\normalsize\it Department of Physical Sciences, College of Science and Engineering,} \\
{\normalsize\it Ritsumeikan University, Shiga 525-8577, Japan}
}

\date{}

\maketitle

\thispagestyle{empty}


\begin{abstract}
\vspace{0.3cm}
\normalsize

The relativistic extension of the classic stellar structure equations is investigated. 
It is pointed out that the Tolman-Oppenheimer-Volkov (TOV) equation with the gradient equation for local gravitational mass can be made complete as a closed set of differential equations by adding that for the Tolman temperature with one equation of state, and the set is proposed as the relativistic hydrostatic structure equations. 
The exact forms of the relativistic Poisson equation and the steady-state heat conduction equation in the curved spacetime are derived.
The application to an ideal gas of particles with the conserved particle number current leads to a strong prediction that the heat capacity ratio almost becomes one in any Newtonian convection zone such as the solar surface. 
The steady-state heat conduction equation is solved exactly in the system and thermodynamic observables exhibit the power law behavior, which implies the possibility for the system to be a new model of stellar corona and a flaw in the earlier one obtained by using the non-relativistic stellar structure equations.
The mixture with another ideal gas yields multilayer structure to a stellar model, in which classic stellar structure equations are reproduced and analytic multilayer structure of luminous stars including the Sun is revealed in suitable approximation.

\end{abstract}
\end{titlepage}

\section{Introduction} 
\label{Introduction}

As also read from descriptions in myths and Bible, the brightness of stars has been recognized as something indispensable and reverent from ancient time. 
The Sun has given energy to living beings during the day, shining stars have attracted people like jewels in the night.
How they maintain energy and beauty has been a mystery before physicists unveil it. 
To reveal the mystery, Helmholtz and Kelvin proposed a hypothesis that the brightness of the Sun originated in the gravitational potential energy due to the contraction of solar fluid
\cite{HelmholtzLXIVOT,kelvin1889popular}. (See \cite{Bahcall:2000xc} for related history.)
This proposal was rejected because only the gravitational contraction cannot produce sufficient energy to maintain the solar visibility longer than the time-scale of terrestrial minerals.
Taking into account the progress of nuclear physics, 
Eddington pointed out that what is necessary for the Sun to gain sufficient luminous energy is certain subatomic reaction process \cite{eddingtoninternal}, whose insight was subsequently developed further by Gamov and Bethe \cite{PhysRev.53.595,Bethe:1939bt}.
For a spherically symmetric star, this Eddington's insight is described by the following equation 
\beal{
{dL_r\over dr}=& 4\pi r^2\varrho \epsilon ,
\label{Lr'}
}
where $L_r$ is the energy flowing outwards across a sphere of radius $r$ called luminosity, $\varrho$ the mass density, and $\epsilon$ the net energy production rate per unit mass liberated from all reaction processes including subatomic ones. If such produced energy is transported by radiation, then the temperature gradient inside a star is induced as \cite{eddingtoninternal}
\beal{
{dT\over dr}
=& -\frac{3\kappa \varrho }{16\sigma T^3}  \frac{L_r}{4\pi r^2} ,
\label{T'Radiation}
}
where $\sigma=\pi^2 ck_{\text B}^4/(60(c\hbar)^3)$ is the Stefan-Boltzmann constant and $\kappa$ is the opacity of stellar fluid per unit mass, while if energy is transported by convection, the temperature gradient is replaced by
\beal{
{dT\over dr}
=& -(1 - \frac1\gamma )\frac{G_{\rm N} M_r \varrho T}{r^2P},
\label{T'convection}
}
where $\gamma$ is the specific heat ratio of stellar fluid, $P$ the local pressure, $G_{\rm N}$ the Newton constant, $M_r$ defined by 
\beal{
{dM_r \over dr}=&4\pi r^2 \varrho. 
\label{Mr'}
}  
On the other hand, the visibility of a luminous star is based on its stable existence, which is achieved as a result of the balance between the gravitational force and the internal pressure of stellar fluid.
In a region where the gravitational force can be approximated by the Newtonian one, the hydrostatic equilibrium can be described by 
\beal{
{dP\over dr}=& -\frac{G_{\rm N} M_r}{r^2}\varrho.
\label{P'}
}

The above set of differential equations is basic to investigate stellar structure and known as the stellar structure equations. 
Solving them combining equations of state for $P,\epsilon,\kappa$ in terms of $\varrho,T$ with suitable boundary conditions, one can determine four macroscopic variables $\varrho, T, M_r, L_r$ as functions of $r$. (See standard textbooks \cite{harwit1988astrophysical,phillips1994physics,tayler1994stars,rnaasaaaa67bib2,2013sse..book.....K,choudhuri2010astrophysics} and references therein.)
These differential equations also yield information on interdependent relations between global observables of luminous stars such as the mass-luminosity relation \cite{1938ApJ....88..472K} and the luminosity-temperature one depicted by the Hertzsprung-Russell diagram \cite{PhysRev.53.595}. 
(See \cite{rnaasaaaa67bib5,Cuntz_2018} for their recent data.)

The above traditional stellar structure equations have certainly played important roles to extract information on interior of luminous stars and interdependent relations of their global observables. However, they are apparently built on the basis of non-relativistic physics, which restricts them to be applicable only in the Newtonian regime such that density and pressure are sufficiently small like in the neighborhood of a stellar surface and the inside of a light star. (See \cite{Misner:1974qy} for instance.)
Indeed, the relativistic extension of the equation for hydrostatic equilibrium \eqref{P'} has already been investigated and established as the Tolman-Oppenheimer-Volkov (TOV) equation \cite{PhysRev.55.374}, and the deviation between non-relativistic results and relativistic ones increases as the mass density does \cite{osti_4092046,1964SvA.....8..147S}. (See also \cite{hartle2003gravity}.)
Since there is an evolutionary process from a main sequence star to a highly dense one consisting of degenerate matter \cite{10.1093/mnras/87.2.114}, it would be desirable to extend the stellar structure equations fully to their relativistic ones so as to be applicable in the non-Newtonian regime such as the interior of a degenerate star and in order to keep track of a stellar evolutionary process.

With taking this into account, it is natural to ask how the traditional structure equations should be extended to the general relativistic ones, and whether any significant consequence can be drawn from this extension particularly to physics of luminous stars less dense than compact stars. (See textbooks, for instance \cite{zel2014stars}, on the study of a compact star using the TOV equation.)
The purpose of this paper is to address these questions by employing some latest results of the author on relativistic local thermodynamics in relativistic hydrostatic equilibrium with spherical symmetry \cite{Yokoyama:2023nld}. 
In the work, entropy current and entropy density were constructed by the method proposed in \cite{Aoki:2020nzm}, and the constructed entropy density was shown to satisfy the local Euler's relation and the first law of thermodynamics concurrently and non-perturbatively with respect to the Newton constant, in which the local temperature is exactly coincident with the one given by Tolman \cite{PhysRev.35.904}, which is called the Tolman temperature in this paper.
On top of this, the established local thermodynamics was applied to a hydrostatic equilibrium system with uniform energy density and the relativistic stellar structure was completely determined with the analytic expressions of all local thermodynamic observables. 
A lessen derived from this simple application is that the TOV equation and two gradient equations for local gravitational mass and the Tolman temperature form a closed set of differential equations with one equation of state given.

Based on these results, the author puts forward that the TOV equation and two gradient equations for the gravitational mass and the Tolman temperature are the relativistic structure equations for spherically symmetric hydrostatic equilibrium upgraded from the traditional stellar structure equations. 
A nontrivial point in this proposal is that the proposed relativistic structure equations do not contain variables on rate such as luminosity and energy production rate, so that they cannot be obtained simply by extending each traditional non-relativistic stellar structure equation to the relativistic one.
Then the main issue of this paper is twofold: whether the proposed structure equations can reproduce the classic stellar structure equations leading to important stellar properties, and whether any new significant consequence can be drawn from the proposed relativistic ones properly. 

The goal of the paper is to answer these questions positively and provide evidence for validity and efficiency of the proposal. 
This is not trivial at all taking into account the fact that the proposed relativistic hydrostatic structure equations have less number of equations and that of variables as well. 
To the end, firstly, the setup of a relativistic hydrostatic equilibrium system with spherical symmetry is fixed and the necessary and sufficient set of the proposed structure equations is explicitly presented in section \ref{StructureEquations}. 
It is pointed out that the proposed temperature gradient equation is consistent with a thermodynamic relation known in the ordinary thermodynamics and thus it is expected to hold for any local thermodynamic equilibrium system. 
Then the relativistic Poisson equation is derived non-perturbatively in the Newton constant, and it is converted into the differential equation for the Tolman temperature using its relation to the gravitational potential.
This is the steady-state heat conduction equation exactly holding in this curved spacetime and plays a key role to determining the hydrostatic structure. 

In section \ref{Applications}, the proposed hydrostatic structure equations are applied to the construction of a model with multilayer structure of luminous stars including the Sun. Through this application, advantages of the relativistic extension of the structure equations become transparent. 
One of the advantages is seen in the temperature gradient equation. In the conventional stellar structure equations, the temperature gradient equation needs to be chosen as either \eqref{T'Radiation} or \eqref{T'convection} suitably in accordance with the way of energy transportation, or is newly computed by using the so-called mixing length theory \cite{1948ZA.....25..135B,vitense1953wasserstoffkonvektionszone}.
In the proposed ones, the temperature gradient equation is unchanged regardless of transport phenomena and the energy transport of fluid is described by its equation of state. 
In particular, it is shown that a simple fluid whose pressure is proportional to its energy density satisfies the conventional temperature gradient equation in a convective zone \eqref{T'convection}. 
Such a simple fluid can be realized by an ideal gas of particles with conserved particle number current, which is called baryonic particles in this paper. This model predicts that the heat capacity ratio is almost one in any Newtonian convection zone such as the solar surface. 
A characteristic feature of this new hydrostatic equilibrium model is that the steady-state heat conduction equation can be solved exactly and thermodynamic observables are determined non-perturbatively in the Newton constant. They exhibit the power law behavior with the power law index related to the heat capacity ratio and imply that this model can be also used to describe an ionized state of fluid and a plasma one in stellar corona. An interesting feature in this model is that pressure is totally well-behaved and vanishing at the asymptotically far region. This result conflicts with the earlier one obtained by using the non-relativistic Newtonian gravity \cite{1958ApJ...128..664P}, and it concludes that the Newtonian approximation is not valid any more in stellar corona. 
After applying to an ideal gas of non-relativistic particles to determine the local temperature by perturbation in section \ref{NRParticle}, an analytic stellar model is investigated as the hydrostatic equilibrium of two types of ideal gases of baryonic particles and non-relativistic ones in section \ref{PhotonBaryon}. In the model, the ideal gas of baryonic particles forms a layer of the main stellar material and that of non-relativistic ones does a layer of atmosphere. 
In addition, by considering a situation with the system coupling to radiation, the luminosity can be related to the local temperature, which leads to the relativistic extension of the conventional temperature gradient equation in a radiation zone \eqref{T'Radiation}.
A summary of the multilayer stellar model is given in table \ref{StellarModel}. 
\begin{table}[thb]
 \begin{center}
  \begin{tabular}{|c|c|c|c|c|}
  \hline
  \hline
Layer & Region & Temperature $T$ & Energy density $\rho$ & Matter \\ 
  \hline
Empty & $ R_{\star} \leq r <\infty $ & $\displaystyle{T_\star \sqrt{\frac{1 - \frac{r_{\rm H}}{R_{\star}} }{1 -\frac{r_{\rm H}}{r} }}}$ &$0$ & ($\gamma$) \\ 
  \hline
${ \mbox{Atmosphere} }$& $R_{*} \leq r \leq R_{\star} $ & $\displaystyle{ T_\star\left(1 -\frac{G_{\rm N}\nu_{*}}{2c^2} +\cdots \right)}$ &$( \bar m c^2 + \frac32 k_{\text B}T)\hat n_{\text g}$ & NR(+$\gamma$) \\ 
  \hline
${\mbox{Interior}}$& ${R_{\rm c} \leq r \leq R_{*} }$ & $\displaystyle{ T(R_{*})\left(\frac{R_{*}}{r}\right)^{2(1 - \frac{1}{\gamma_{\rm b}}) } }$  &$\displaystyle{ \rho(R_{*})\(\frac{R_*}r\right)^2 }$ & NR+B(+$\gamma$) \\ 
  \hline
Core & $ 0 < r \leq R_{\rm c} $ & $ T(R_{\rm c}) \frac{ 2 \sqrt{r_{\rm c}^2-R_{\rm c}^2} }{3\sqrt{r_{\rm c}^2-R_{\rm c}^2} -  \sqrt{r_{\rm c}^2-r^2} } $ &$\rho_{\rm c}$ & D \\ 
  \hline
  \hline
  \end{tabular}
 \caption{A summary of a simple analytic multi-layer stellar model is shown. In each layer, the local temperature and the energy density determined in the main text are shown. In the column of 'Matter', the symbols 'NR', 'B', '$\gamma$', and 'D' mean 'Non-Relativistic particles', 'Baryonic particles', 'photons', and 'Degenerate matter', respectively. The reason why photon is enclosed by bracket is that it is not involved in the stellar structure equations. 
 $r_{\rm H}= \frac{2G_{\rm N}M_\star}{c^2} , r_{\rm c} =\sqrt \frac{3c^2}{ 8\pi G_{\rm N} \rho_{\rm c} }$, $M_\star$ is the stellar gravitational mass, $\gamma_{\rm b}$ is the heat capacity ratio of the ideal gas of baryonic particles, while $\nu_*$ is given by \eqref{nu*}, $\hat n_{\text g}$ \eqref{ng}, $T_\star$ \eqref{SB}, $R_\star$ \eqref{Rstar}, $T(R_*)$ \eqref{T*Tstar}, $R_*$ \eqref{R*}, in the main text.}
 \label{StellarModel}
 \end{center}
\end{table}
Employing obtained results, a concrete model of the Sun is constructed in section \ref{SolarModel}.

Section \ref{Discussion} is devoted to summary and discussion including open problems and future works.

\section{Relativistic hydrostatic structure equations} 
\label{StructureEquations}

In order to generalize the hydrostatic structure equations so as to include relativistic effect, a hydrostatic equilibrium system needs to be formulated by general theory of relativity. 
Such a system with spherical symmetry formulated by general relativity was studied in \cite{PhysRev.35.904,PhysRev.55.374,1935ApJ....82..435H}, (see also appendix in \cite{Hawking:1973uf}) in which the line element generally takes the form such that 
\beal{
g_{\mu\nu}\mathrm dx^\mu \mathrm dx^\nu
=& - e^\nu \mathrm (cdt)^2 +e^\lambda \mathrm dr^2 + r^2(\mathrm d\theta^2+(\cos\theta)^2\mathrm d\phi^2),
\label{metric}
}
where $\nu, \lambda$ are functions of the radial coordinate $r$, 
and the energy-stress tensor is of the form of a perfect fluid 
\beal{
T_{\mu\nu} = (\rho+P)\frac{u_\mu}c \frac{u_\nu}c + P g_{\mu\nu},
}
where $\rho$ is local energy density, $P$ is local pressure, $u^\mu$ is the fluid four velocity. 
Note that $c^2\nu/2$ corresponds to the gravitational potential probed by a point particle with unit mass in the weak gravity regime \cite{zel2014stars}, so $\nu$ itself will be also called the gravitational potential for convenience. 
The balance between the gravitational attractive force and the repulsive one of fluid is described by the Einstein equation, which reduces in the comoving frame to 
\beal{
P'
=& -\frac{(P+\rho)}{2}\nu',
\label{CCEr3} \\
\lambda'=& r e^\lambda 8\pi G\frac\rho{c^2} -\frac{e^\lambda-1}{r},
\label{EOM1} \\
\nu'=& r e^\lambda 8\pi G \frac P{c^2} + \frac{e^\lambda-1}{r},
\label{EOM2}
}
where $G:=G_{\rm N}/c^2$. 
Eliminating $\lambda$ and $\nu$ leads to the TOV equation \cite{PhysRev.55.374}
\beal{
P'
=& -\frac{(P+\rho)}{c^2} \frac{ G( \check E_r+4\pi r^3 P)}{ r^2(1-2\frac{ G\check E_r }{rc^2}) }, 
\label{TOV} \\
\check E_r'=& 4\pi r^2\rho,
\label{rho}
}
where $\check E_r$ is defined by $\check E_r := {rc^2\over 2 G}(1-e^{-\lambda} )$. Note that $\check E_r$ has the dimension of energy. 

Employing the method given in \cite{Aoki:2020nzm} one can construct entropy current and entropy density as a conserved current and a conserved charge density for this system \cite{Yokoyama:2023nld}. (See also \cite{Aoki:2020prb}.)
A key to the construction is to find a vector field $\xi^\mu$ to satisfy a differential equation such that $T^\mu\!_\nu \nabla_\mu \xi^\nu=0$. This vector field enables one to construct a conserved current as $J^\mu=\sqrt{-g}T^\mu\!_\nu \xi^\nu$ and a conserved charge as $Q= \int d^3x J^t$ in general. In order to find conserved entropy current for the fluid, look for such a vector field $\xi^\mu$ to be proportional to the fluid velocity $u^\mu$. The entropy density constructed in this way was shown to satisfy the local Euler's relation and the first law of thermodynamics concurrently and non-perturbatively with respect to the Newton constant in the comoving frame \cite{Yokoyama:2023nld}. 
\be
Ts = u+Pv, ~~ 
Ts' = u'+Pv', 
\label{Eular1stLaw}
\ee
where $v=r^2 \cos\theta(1 - {2 G\check E_r \over rc^2})^{-\half}$, $u=\rho v$ is the internal energy density, $T\propto e^{-\frac\nu2}$ is the local temperature given by Tolman \cite{PhysRev.35.904}. 
In particular, it was shown that the entropy density does not depend on the radial coordinate, which is a natural consequence for a steady state with vanishing heat flux in the energy stress tensor.
Using \eqref{CCEr3}, one can derive the temperature gradient equation as \cite{Yokoyama:2023nld}
\be 
T' = {P' \over P+\rho} T. 
\label{Tgradient0}
\ee
Plugging \eqref{TOV} into this leads to 
\be 
T' = - \frac{ G( \check E_r+4\pi r^3 P)}{c^2 r^2(1-2\frac{ G\check E_r }{rc^2}) } T. 
\label{Tgradient}
\ee
This temperature gradient equation \eqref{Tgradient} and the TOV equation \eqref{TOV} with \eqref{rho} form a closed set of differential equations including one equation of state, and this set of three differential equations is proposed as the relativistic extension of classic stellar structure equations. 

Comments are in order. 
Firstly, the temperature gradient equation \eqref{Tgradient0} involves only thermodynamic variables and their gradients, which suggests that a corresponding relation exists in global thermodynamics on flat spacetime. The answer of such a thermodynamic relation is
\be 
\bar T\({\partial\bar P\over \partial\bar T}\)_{\bar V,\bar N}
=\({\partial\bar U\over \partial\bar V}\)_{\bar T,\bar N}+\bar P, 
\label{Maxwell}
\ee
since this can be rewritten as $({\partial\bar P\over \partial\bar T})_{\bar V,\bar N}=\frac{\bar T}{\bar P+\bar\rho}$, where $\bar\rho=({\partial\bar U\over \partial\bar V})_{\bar T,\bar N}$ and the thermodynamic variables with overline stand for global quantities used here only.%
\footnote{ 
It may not be so trivial to obtain the local thermodynamic relation \eqref{Tgradient0} from \eqref{Maxwell}, because the energy density $\rho$ appearing in the energy stress tensor and the internal energy density $u$ in the thermodynamic relations are different on curved spacetime as shown above, which could make it difficult to presume \eqref{Tgradient0} from \eqref{Maxwell}.
}
Therefore the relation \eqref{Tgradient0} is expected to hold for any local thermodynamic equilibrium system. 
Secondly, as is clear from the above derivation, there are several equivalent expressions for the temperature gradient equation. \eqref{Tgradient} may be useful for numerical analysis, while \eqref{Tgradient0} or \eqref{CCEr3} may be more useful for non-numerical one. Which expression is best to be used will depend on contexts and ways of analysis. 
Thirdly, although the constructed entropy density does not play any role in the structure equations, the existence thereof has important implications. One is that this ensures the validity of the definitions of the macroscopic quantities such as the temperature and the internal energy due to the fact that they satisfy the local Euler's relation and the first law of thermodynamics. Another is to assure the system to be in a local equilibrium, which justifies to solve the hydrostatic structure equations by assuming the distribution of an ideal statistical ensumble for local thermodynamic quantities. 
This point together with \eqref{Tgradient0} will become important to include the temperature effect dynamically. 
In earlier study, the TOV equation \eqref{TOV} and \eqref{rho} were used to determine the behavior of pressure and density by solving them combined with an equation of state with respect to pressure and density only \cite{zel2014stars}, but they can be used to determine the local temperature and include the thermal effect incorporating the temperature gradient equation \eqref{Tgradient0}. 
Fourthly, the proposed hydrostatic structure equations do not contain the rate variables such as the luminosity or the corresponding equation for the luminosity gradient \eqref{Lr'}. 
They are closed without the variable of luminosity, which implies that the variables on rate are not fundamental for hydrostatic equilibrium. This result is preferable to the fact of the existence of a non-radiating star in hydrostatic equilibrium, while this in turns implies that the proposed hydrostatic structure equations are not applicable to a situation where there is so strong radiant energy flux to involve hydrostatic structure. The restriction of applicability is discussed in section \ref{EddingtonLimit}.
Finally, on the non-relativistic reduction, it is easily seen that the TOV equation with \eqref{rho} reduce to the non-relativistic hydrostatic equation \eqref{P'} with \eqref{Mr'} in the Newtonian regime where $P\ll\rho, P\ll\frac{\check E_r}{4\pi r^3},\frac{2G_{\rm N}\check E_r}{c^2} \ll r$ so that $\rho\to\varrho c^2, \check E_r\to M_r c^2$. On the other hand, it needs some preparation of setup to reproduce \eqref{T'Radiation} and \eqref{T'convection}, so it will be confirmed later. 

\subsection{General relativistic Poisson equation and steady-state heat conduction equation} 
\label{StellarTemperature} 

Before moving on to detailed applications of the proposed relativistic hydrostatic structure equations, it is convenient to investigate how to determine local temperature from them. 
From the definition of $\check E_r$ and \eqref{EOM2}, a concise expression of $\check E_r$ with respect to thermodynamic variables is derived as 
\be 
\check E_r=\frac{\frac{c^2 r^2 \nu'}{G}-8 \pi  r^3 P}{2 (1+ r \nu')}, 
\label{Er}
\ee
with $\nu'= -2T'/T$.%
\footnote{ 
The leading expression in \eqref{Er}, $\check E_r\approx\frac{c^2 r^2 \nu'}{2G}$, may be reminiscent of the conventional mass expression of a star derived by incorporating the polytropic relation $P\propto\varrho^{1+1/n}$ and the Lane-Emden equation $\frac1{\xi^2}{d\over d\xi}(\xi^2{d\theta\over d\xi})=-\theta^n$ with $\xi\propto r$ as $m \propto \xi_1^2 |\theta(\xi_1)'|$, where $\theta(\xi_1)=0$ \cite{1935MNRAS..95..207C}. 
However, the expression \eqref{Er} is much more general in the sense to hold without imposing any equation of state and to include the effects not only of general relativity but also of nonzero temperature. 
}
Differentiating both sides with respect to $r$ and using \eqref{CCEr3} and \eqref{rho}, one finds 
\beal{
\nu''+\frac{2 \nu'}{r}+\nu'^2-
\frac{4 \pi  G }{c^2}\((r \nu'+1) (r \nu'+2) \rho- \left(2 r^2 \nu''+r\nu' (r \nu'-3)-6\right)P\)=0. 
\label{GRPoisson}
}
This can be rewritten into a general coordinate invariant form as
\beal{
\nabla^2\nu = \frac{8\pi G}{c^2} (g^{\omega\mu}+2u^\omega u^\mu)T_{\omega\mu}, 
\label{GRPoisson0}
}
which is the general relativistic extension of the Poisson equation. 
To prove this, first compute the Laplacian acting on the scalar field $\nu$ as 
\beal{
\nabla^2 \nu 
=&e^{-\lambda}(\nu'' + (\frac2r + \frac{\nu'-\lambda'}2 )\nu' ) \nn
=&e^{-\lambda}\nu'' +\frac{  \nu'+4 \pi  G r/c^2 (P (r \nu'+3)-(r \nu'+1) \rho )+2/r }{r \nu'+1}\nu',
\label{boxnu}
}
where at the 1st equation the formula $\nabla^2 \nu =\frac1{ \sqrt{|g|}}\partial_\omega \sqrt{|g|} g^{\omega\mu} \partial_\mu \nu$ was used and at the 2nd one the term containing $\lambda'$ was computed by employing \eqref{EOM1} and \eqref{EOM2} as 
\beal{
e^{-\lambda} \frac{\nu'-\lambda'}2
= 4\pi G r\frac {P-\rho}{c^2} + \frac{2 G\check E_r}{r^2c^2}, 
}
and \eqref{Er} was used to substitute $\check E_r$.
The coefficient of $\nu''$ in \eqref{boxnu} is $e^{-\lambda}=\frac{1+8 \pi  G r^2 P/c^2}{r \nu'+1}$, while that in \eqref{GRPoisson} is $1+8 \pi  G r^2 P/c^2=e^{-\lambda}(r \nu'+1)$. 
Thus multiplying $r\nu'+1$ for both sides in \eqref{GRPoisson} and subtracting it from \eqref{boxnu}, one can remove $\nu''$ in \eqref{boxnu} and the rest is \cite{zel2014stars}
\be 
\nabla^2 \nu =8 \pi  G (\rho + 3 P )/c^2.
\label{GRPoisson1}
\ee
This matches \eqref{GRPoisson0} and completes the proof. 
It is important to stress that the form of the relativistic Poisson equation \eqref{GRPoisson0} or \eqref{GRPoisson1} has been derived here without using any approximation and holds non-perturbatively in the Newton constant.

The generalized relativistic Poisson equation can be solved order by order in the Newton constant. The leading order is 
\be 
\nu''+\frac{2 \nu'}{r}+\nu'^2= 0. 
\label{nu0} 
\ee 
This has a trivial solution that $\nu$ is constant, while there is a non-trivial one as $\nu'= \frac1{r(-1+C_0r)}=:\nu_\star'$, 
where $C_0$ is an integration constant.
This non-trivial solution corresponds to the exact one of the original differential equation \eqref{GRPoisson} in the matter-empty region, $P=\rho=0$. 
The integration constant $C_0$ is fixed by plugging this back into \eqref{Er}
\beal{
\check E_r = \frac{c^2-8 \pi  G r^2 (-1+C_0 r) P}{2 C_0 G}.
}
Then evaluate this at a point $r=R_\star$ so far away from the center of the system that the contribution of the pressure to $\check E_r$ is negligible, $\check E_r\approx\frac{c^2 }{2 C_0 G}$ with $r\geq R_\star$. In this far region, $\check E_r$ is approximated as a constant and known as the gravitational mass by dividing it by $c^2$, which is denoted by $M_\star$. Then the integration constant is fixed as $C_0 = \frac{1}{2 M_\star G}$ and $\nu_\star$ is given by $\nu_\star=\log \left(\frac1r-\frac1{2 M_\star G}\right)+\tilde C_0$, where $\tilde C_0$ is another integration constant. 
This integration constant originates in the arbitrariness of the base point of the gravitational potential, which can be fixed by choosing an arbitrary referencing point. Here it is fixed by requesting it to vanish at the evaluation point: $\nu(R_\star)=0$. Then the gravitational potential in the matter-empty region is determined as 
\be 
\nu_\star=\log \left(\frac{1 -\frac{2 M_\star G}{R_\star} }{1 -\frac{2 M_\star G}r }\right). 
\label{nuStar}
\ee
This exact gravitational potential in the matter-empty region is expanded with respect to the Newton constant as 
\be 
\nu_\star = 2 M_\star G \( -\frac1r+\frac1{R_\star} \)+\cdots . 
\label{nuStar1}
\ee
This leading term describes the Newton's law of universal gravitation for an object of the mass $M_\star$, which is the origin of the name of $M_\star$ mentioned above. 
As a result the evaluation point $r=R_\star$ should be chosen to satisfy the pressure at the point $P(R_\star)$ to satisfy an inequality 
\be 
c^2\gg 4\pi R_\star^3 \frac {P(R_\star)}{M_\star} ,
\label{radiusCondition}
\ee
with $R_\star\gg r_{\rm H}$ assumed. 
Accordingly the behavior of temperature in the matter-empty region is exactly determined as \cite{Yokoyama:2023nld}
\beal{
T=T(R_\star) e^{-\frac{\nu}2}= T(R_\star) \sqrt{\frac{1 -\frac{r_{\rm H}}{R_\star} }{1 -\frac{r_{\rm H}}r }}=:T_{\rm vac},
\label{Tstar} 
}
where $r_{\rm H}:={2 M_\star G}$ is the radius of the horizon of the (Schwarzschild) black hole whose mass is identical to the stellar gravitational mass. 
The local temperature is formally divergent at the location of the horizon radius. 
It can be confirmed that \eqref{Tstar} satisfies the differential equation $T_{\rm vac} \left(T_{\rm vac}''+\frac2r T_{\rm vac}'\right)-3 T_{\rm vac}'^2 =0$, which is identical to the leading order of the relativistic Poisson equation written in term of the local temperature as 
\beal{
&T \left(T''+\frac2r T'\right)-3 T'^2  \nn
=&\frac{4 \pi  G}{c^2} \(\rho  \left(-2 r^2 T'^2+3 r T T'-T^2\right)+ P \left(4 r^2 T'^2-r T \left(2 r T''-3 T'\right)-3 T^2\right)\) . 
\label{GRPoissonT}
}
This is the relativistic steady-state heat conduction equation for a relativistic hydrostatic equilibrium system with rotational symmetry.  

The next-to-leading order of the general relativistic Poisson equation in the Newton constant is determined by expanding the differential equation \eqref{GRPoisson} as $\nu=\nu_0+G \nu_1+\cO(G^2)$ with $\nu_0$ constant and taking the leading term as
\be 
\nu_1''+\frac 2r \nu_1'=\frac{8 \pi }{c^2}(\rho_0 +3P_0), 
\label{PoissonNLO} 
\ee
where $\rho_0,P_0$ are the leading order of the energy density and that of the pressure, respectively. 
In terms of the variable of temperature, the next-to-leading order of the relativistic Poisson equation is given by 
\be 
T_1''+\frac 2r T_1'=-\frac{4 \pi T_0}{c^2}(\rho_0 +3P_0), 
\label{PoissonNLOT} 
\ee
where $T=T_0+GT_1+\cdots$ with $T_0$ constant.
Inputing detailed information on fluid and solving the differential equation \eqref{PoissonNLO}, \eqref{PoissonNLOT} gives the subleading correction of the gravitational potential, the local temperature, respectively.

Comments are given below. 
Firstly, a perturbative solution for \eqref{GRPoisson} or \eqref{GRPoissonT} obtained this way is useful in a weak-gravity regime with energy density and pressure sufficiently small such as the vicinity of a stellar surface, while in a strong-gravity region such as the interior of a heavy stellar core, more useful information will be obtained by solving the original relativistic Poisson equation directly in a non-perturbative fashion with respect to the Newton constant. 
Secondly, astrophysical observation is basically done in a region far away from the core, so $r\gg r_{\rm H}$. 
Therefore, the condition $r \gg 2G\check E_r/c^2 $ is assumed for practical applications in what follows.
In particular, this condition is satisfied in the Newtonian regime mentioned above. 

\section{Applications} 
\label{Applications}

In this section, the relativistic hydrostatic structure equations proposed in the previous section are applied to constructing a model of luminous stars. This application also demonstrates how classic results on a luminous star are reproduced or modified with taking into account the effect of general relativity and how to determine local thermodynamic observables inside a luminous star.

\subsection{Ideal gas of baryonic particles} 
\label{IF}

To the end of the construction of a model of luminous stars, first consider a hydrostatic equilibrium system of a single ideal gas of particles with conserved particle number current, which are called baryonic particles in this paper. 
The equation of state of an ideal gas is 
\be 
P =\hat n k_{\text B} T, 
\label{EOSidealgas}
\ee
where $\hat nv$ gives the ordinary particle number density and $k_{\text B}$ is the Boltzmann constant, while the conservation of the particle number current requires that $\nabla_\mu J^\mu = 0$, where $J^\mu=\hat n u^\mu$. The left hand side is computed in a radially moving frame as $\nabla_\mu J^\mu = u^r(\hat n' + \hat n (\log(\sqrt{|g|}u^r))' )$. 
A relativistic fluid equation derived in \cite{Yokoyama:2023nld} rewrites this as $\nabla_\mu J^\mu = u^r(\hat n' - \hat n\frac {\rho'}{\rho + P}  )$. 
Therefore the conservation of the number current leads to 
\be 
\hat n' = \hat n\frac {\rho'}{\rho + P} . 
\label{Continuity}
\ee
Substituting \eqref{EOSidealgas} into this so as to eliminate $\hat n$ and using \eqref{Tgradient0}, one finds $P'/P=\rho'/\rho$. This can be solved as
\be 
P=w\rho,
\label{EOS}
\ee
where $w$ is an integration constant. 
The fluid for pressure to be proportional to energy density such as \eqref{EOS} is called a simple fluid in what follows for convenience. 
To fix the integration constant, recall that the energy density of an ideal gas is given by $u = n C_V T$, where $C_V$ is the heat capacity at constant volume and $n=\hat nv$. This can be rewritten as $\rho =\hat nC_VT$. By substituting this and \eqref{EOSidealgas} into \eqref{EOS}, the integration constant can be fixed as  
\be
w = \frac{k_{\text B}}{C_V} = \gamma - 1, 
\label{w2HCR}
\ee
where $\gamma=C_P/C_V$ is the heat capacity ratio.
Here $C_P$ is the heat capacity at constant pressure and given by the Mayer's relation as $C_P=C_V+k_{\text B}$, which is guaranteed to hold by the laws of thermodynamics confirmed above. 
Plugging \eqref{EOS} into \eqref{Tgradient0} with \eqref{w2HCR}, one obtains 
\be 
\frac{T'}T =(1-\frac 1{\gamma}) {P' \over P}. 
\label{Convection}
\ee
This is exactly the saturation point of the inequality of the Schwarzschild stability condition for a hydrostatic equilibrium system \cite{1906Schwarzschild} and known as an equation of state characteristic to a convection zone, in which the energy transport occurs mainly by convection.
On the other hand, in order for fluid in hydrostatic equilibrium to satisfy \eqref{Convection}, it needs to be a simple fluid satisfying \eqref{EOS}. As a result, the necessary and sufficient condition for fluid to satisfy the equation \eqref{Convection} is that there is a proportionality relationship between pressure and energy density of fluid in the region. 
Note that the equation \eqref{Convection} can be easily solved as $P \propto T^{\frac\gamma{\gamma-1}}. $ Thus $\rho = aT^{\frac\gamma{\gamma-1}}$, where $a$ is a constant. 

This new derivation of the temperature gradient equation in the convection zone leads to a new important prediction for a general property of stars. 
That is, in a region with the Newtonian approximation valid, pressure is negligible compared to energy density so that the parameter $w$ almost vanishes \cite{Misner:1974qy}.
Combining this with the observational fact that the Newtonian approximation is valid near the surface of most stars, one concludes that the heat capacity ratio in any stellar convection zone with the Newtonian approximation valid is almost one:%
\footnote{ 
This is consistent with a result in fluid dynamics on flat spacetime that an isentropic fluid satisfies a polytropic relation $P\propto \rho^\gamma$. 
}
\be 
\gamma \sim 1. 
\label{StellarHCR}
\ee
In particular, it is known that there exists considerably large convection zone underneath the solar surface \cite{RevModPhys.60.297,NASASolarPhysics}. Therefore, it follows that fluid near the solar surface has the heat capacity ratio nearly equal to $1$. 
This is a robust prediction newly made in this paper.
The deviation of the heat capacity ratio from one will be described by stellar global observables as \eqref{HCR*} later. 

\subsubsection{Power law behavior of thermodynamic observables}
\label{PowerLaw} 

The radial dependence of temperature is determined by solving the relativistic steady-state heat conduction equation 
\eqref{GRPoissonT}, which reduces to
\beal{
&T \left(T''+\frac2r T'\right)-3 T'^2  \nn
=&\frac{4 \pi a G}{c^2} T^{\frac\gamma{\gamma-1}} \(-6 r^2 T'^2+2 r^2 T T''+2T^2+ \gamma \left(4 r^2 T'^2-r T \left(2 r T''-3 T'\right)-3 T^2\right)\) . 
\label{GRPoissonTPF}
}
This can be solved perturbatively in the Newton constant as illustrated in section \ref{StellarTemperature}, while this expression implies the existence of an exact solution obeying the power law, $T=A r^n$, where $A$ is a nonzero constant and $n$ is a negative number so that the temperature increases as it goes deeper into the center. 
These parameters are determined by substituting back into the differential equation, which simplifies to 
\be 
A^2 (2 n-1) r^{2 n-2} \left(4 \pi  a G A^{\frac\gamma{\gamma-1}} (n (\gamma-2)+3 \gamma-2 ) r^{\frac{n}{\gamma-1}+n+2}+c^2 n\right) = 0. 
\ee
There exists a nontrivial solution if and only if $\frac{n}{\gamma-1}+n+2=4 \pi  a G A^{\frac\gamma{\gamma-1}} (n (\gamma-2)+3 \gamma-2 ) r^{\frac{n}{\gamma-1}+n+2}+c^2 n=0$. Thus the parameters for a nontrivial solution are determined as $n=-2(\gamma-1)/\gamma, A= \left(\frac{c^2 (\gamma-1)}{2 \pi  a G \left(\gamma^2+4\gamma -4\right)}\right)^{\frac{\gamma-1}{\gamma}}$. 
As a result, the relativistic steady-state heat conduction equation has been solved exactly as 
\be 
T 
= \left(\frac{c^2 (\gamma-1)}{2 \pi  a G \left(\gamma^2+4\gamma -4\right)}\right)^{1 - \frac{1}{\gamma}} \frac1{r^{2(1 - \frac{1}{\gamma}) } }.
\label{TIdealFluid}
\ee
That is, the local temperature obeys the power law. 
Thus the density and the pressure do so as well in such a way that $P\propto \rho \propto 1/r^2$. 
Therefore it concludes that a hydrostatic equilibrium system of a simple fluid exhibits the power law behavior of macroscopic observables. 

It is important to stress that this solution cannot be obtained by perturbation in the Newton constant, so that the power law behavior cannot be seen by solving the system perturbatively. 
It is also interesting that the parameters in this solution are given only by data of internal stellar fluid and not by external data such as the gravitational mass.
From this result, the gravitational potential for this model can be also determined as $\nu=4 (1-\frac1\gamma) \log (r/R_\star).$

\subsubsection{New model for stellar corona}
\label{StellarCorona}

The above results of the power law behavior of thermodynamic observables imply that this model is applicable to stellar corona, in which temperature is so much higher than the other part of stellar superficial region that gas is almost or completely ionized. 
This suggests that macroscopic behavior and transport phenomena in stellar corona can be investigated by combining plasma physics with stellar structure equations. For instance, temperature is predicted to fall off in a power law in solar corona. 
It was pointed out by Parker that combining the result of plasma physics for fully ionized gas with the conventional non-relativistic hydrostatic structure equations leads to non-vanishing pressure at infinity \cite{1958ApJ...128..664P}.
He speculated on this as a signal of the impossibility for gas in the solar corona to reach hydrostatic equilibrium and presumed that the non-vanishing pressure at asymptotic region gives rise to gas streaming outward from the Sun called the solar wind, whose existence was suggested by Biermann earlier \cite{1951ZA.....29..274B,1957Obs....77..109B}.
The solar wind was indeed observed by a spacecraft launched some years later \cite{1960SPhD....5..361G,10.1007/978-94-011-7542-5_7}.

This is a successful interplay between theory and experiment. However,  it is also pointed out that there still remains unsolved issues for the corona and the solar wind by using the traditional stellar structure equations and results of plasma physics \cite{verscharen2019multiscale}. 
In the application of the above relativistic hydrostatic model of a simple fluid to stellar corona, however, there is no problem for the system to stay in hydrostatic equilibrium. 
Indeed, not only the local temperature but also the local pressure behave in the power law as shown above, which leads to pressure vanishing in the asymptotic region.

This result conflicts with the earlier one obtained by using the non-relativistic structure equations explained above.
Then one might wonder which result is indeed correct. 
The answer of the author is that the earlier model contains a flaw in the argument. 
To explain this, one first needs to accept the fact that the relativistic hydrostatic structure equations presented in this paper, which are rigorously derived from the Einstein gravity, are more fundamental than the non-relativistic ones derived from the Newtonian gravity. 
Indeed, it has been confirmed that the relativistic hydrostatic structure equations reduce to the non-relativistic ones by neglecting terms due to the effect of general relativity. 
Such an approximation is valid only in the Newtonian regime, where the pressure is negligible compared to other terms. 
This implies that the Newtonian approximation cannot be justified for a solution such that the pressure is non-vanishing at a far distance. 
More concretely, one of the conditions to justify the non-relativistic reduction, $4\pi r^3 P \ll \check E_r$, is clearly violated for such a situation, and the contribution from pressure in \eqref{TOV} is not negligible at all. 
Parker derived the analytic expression of the number density for an ideal gas with temperature behaving as the power law by solving the non-relativistic hydrostatic equation \eqref{P'}, and found that it becomes divergent at infinity \cite{1958ApJ...128..664P}, while, in the current system of an ideal gas with temperature obeying the power law, the number density has been obtained as 
\be 
\hat n = \frac P{k_{\text B}T} = \frac{a(\gamma-1)}{k_{\text B}} \left(\frac{c^2 (\gamma-1)}{2 \pi  a G \left(\gamma^2+4\gamma -4\right)}\right)^{\frac{1}{\gamma}} \frac1{r^{\frac{2}{\gamma}} }, 
\ee
which also obeys the power law and is vanishing in the asymptotic region. The latter fully contains general relativistic effect. 

As a result it concludes that the Newtonian approximation is not valid  for a hydrostatic equilibrium system in a plasma state such as stellar corona. Note that this conclusion does not deny the existence of the solar wind, whose existence must be accounted for by another mechanism. 
It is an interesting future work to study whether a newly proposed model compatible with relativistic structure equations can demonstrate any property on stellar corona and make prediction on it.

\subsubsection{Comment on application to radiative fluid} 
\label{application2radiation}

One might wonder that this result is applicable also to an ideal gas of photon or the radiation dominant region, which is described by the equation of state \eqref{EOS} with $w=\frac13$, so $P_{\rm rad}=\frac13\rho_{\rm rad}\propto T^4$, $\rho_{\rm rad} = a T^4$. 
The integration constant $a$ is fixed by considering a surface of a layer%
\footnote{ 
This layer may be called photosphere, which is conventionally specified by a single surface at a certain optical depth $\bar\tau=\frac23$ defined by $\bar\tau = \int_R^\infty \kappa\varrho dr$,
where $R$ is a radius of a luminous star \cite{2013sse..book.....K}.  
}
at which the Stefan-Boltzmann's law holds, so that $a=4\sigma/c$.
This leads to $\rho_{\rm rad}= \frac{4\sigma}c T^4,$
which is consistent with the Bose distribution in equilibrium.
Then the behavior of the local temperature would be given by \eqref{TIdealFluid} with the parameters specified above for the radiation dominant region. 
However, there is an important caveat for the application of the relativistic hydrostatic structure equations. 
That is, they were derived in the comoving observer with the perfect fluid, and such an observer does not exist for null fluid. Thus the application of the above results to radiation is not supported strictly speaking, though it may be useful merely as a rough estimation. Note that this caveat is also the case for the non-relativistic stellar structure equations.
The proper analysis to include the effect of radiation to hydrostatic structure is left as future work. 

\subsection{Ideal gas of non-relativistic particles} 
\label{NRParticle}

Next, consider a system of an ideal gas consisting of non-relativistic particles for later use. 
The equation of state is given by $P=P_{\text g}, \rho=\rho_{\text g}$, where  
\beal{
P_{\text g}= \hat n_{\text g} k_{\text B} T, ~~~ 
\rho_{\text g}= ( \bar m c^2 + \frac32 k_{\text B}T)\hat n_{\text g},
\label{EOSidealgasNR}
}
with $ \bar m$ the averaged mass of the non-relativistic particles. This averaged mass is described as  $\bar m = \mu_{\rm e} m_{\rm u}$, where $\mu_{\rm e}$ is the mean molecular weight per free electron and $m_{\rm u}$ is the atomic mass unit defined by $1{\rm g/mol}$. 
Note that $\mu_{\rm e}$ can be described by using the weight fraction of hydrogen denoted by $X$ as $\mu_{\rm e}=2/(1+X)$.
Then consider to satisfy the temperature gradient equation \eqref{Tgradient0}, which is in the current setup given by 
$(P_{\text g}+\rho_{\text g})T'=P'_{\text g} T$. 
Solving this differential equation, one can fix the form of the particle number density as
\be 
\hat n_{\text g}= C_{\text g} T^{3/2} e^{-\frac{ \bar mc^2 }{k_{\text B} T}},
\label{ng}
\ee
where $C_{\text g}$ is an integration constant. 
This result is indeed consistent with results of statistical physics, and the integration constant can be fixed to match the Maxwell distribution as $C_{\text g}= \bar g (\frac{2\pi \bar mk_{\text B}}{ h^2})^{\frac32}$, where $\bar g$ corresponds to the mean internal degrees of freedom of a non-relativistic particle. 

As previously, the radial dependence of temperature is determined by solving the steady-state heat conduction equation.
Differently from the previous case of simple fluid, it is very difficult to solve it exactly, so here it is solved up to  the next-to-leading order. 
As shown in section \ref{StellarTemperature}, the leading order solution for \eqref{nu0} is a constant, $\nu_0=\const$, so is the temperature, $T_0=\const$, and thus the above thermodynamic quantities as well: 
\be 
P_{0}= \hat n_{g0} k_{\text B} T_0, ~~~ 
\rho_{0}= ( \bar m c^2 + \frac32 k_{\text B}T_0)\hat n_{g0},~~~
\hat n_{g0}= \bar g (\frac{2\pi \bar mk_{\text B}}{ h^2})^{\frac32} T_0^{3/2} e^{-\frac{ \bar mc^2 }{k_{\text B} T_0}}.
\label{rho0P0}
\ee
Therefore the right-hand side of the next-to-leading order of the relativistic Poisson equation \eqref{PoissonNLO} becomes a constant, so it can be solved as 
\be 
\nu_1= \frac{4 \pi  r^2 (\rho_0+3 P_0)}{3 c^2}-\frac{C_1}{r}+\tilde C_1,
\label{PoissonIF}
\ee 
where $C_1, \tilde C_1$ are integration constants. Plugging this back into \eqref{Er} leads to 
\be 
\check E_r=\frac{8 \pi  c^2 r^4 \rho_0-3 c^4 C_1 r}{16 \pi  G r^3 (\rho_0+3 P_0)+6 c^2 (r+C_1 G)}.
\ee
Evaluating this at the surface of the inner layer, $r=R_*$, one finds 
\be 
c^2 M_*=\frac{8 \pi  c^2 R_*^4 \rho_0-3 c^4 C_1 R_*}{16 \pi  G R_*^3 (\rho_0+3 P_0)+6 c^2 (R_*+C_1 G)},
\ee
where $M_*=\check E_r(R_*)/c^2.$
From this, the integration constant $C_1$ is fixed as 
\beal{
C_1 
=& \frac{2 R_* M_* \left(8 \pi  G R_*^2 (\rho_0+3 P_0)/c^2+3 \right)-8 \pi  R_*^4 \rho_0/c^2 }{3 R_*-6 G M_*} \nn
\approx& 2 M_* -\frac83 \pi  R_*^3 \rho_0/c^2 , 
}
while $\tilde C_1$ is determined to request $\nu_1$ to vanish at $r=R_\star$ as introduced in section \ref{StellarTemperature}, outside which matter contribution is negligible so that \eqref{radiusCondition} is satisfied:
\be 
\tilde C_1
=\frac{2 M_*}{ R_\star} - \frac{4 \pi  \left(3 P_0 R_\star^3+\rho_0 \left(R_\star^3+2 R_*^3\right)\right)}{3 c^2 R_\star}.
\ee
Substituting this into \eqref{PoissonIF} yields the solution of the gravitational potential at the sub-leading order as $\nu_1\approx\nu_*$, where 
\beal{
\nu_*
=2(\frac 1{R_\star}-\frac1r) \left(M_*+ \frac{6 \pi  P_0 r R_\star (r+R_\star)+2 \pi  \rho_0 \left(r^2 R_\star+r R_\star^2-2 R_*^3\right)}{3 c^2}\right). 
\label{nu*}
}
This solution is valid in an annuls region with $R_* \leq r \leq R_\star$ and should connect smoothly to the one in the matter-empty region \eqref{nuStar1} at $r=R_\star$, which imposes a condition to satisfy a relation between the stellar gravitational mass $M_\star$ and $M_*$ as
\be 
M_\star = M_*+4\pi R_\star^3 \frac{P_0}{c^2}+\frac43\pi \left(R_\star^3- R_*^3\right) \frac{\rho_0 }{c^2},
\label{MstarM*}
\ee
at the leading order of the Newton constant.
Then the local temperature is determined up to the next-to-leading order in $G$ as 
\be 
T=T(R_\star)(1 - G\nu_*/2+\cdots),
\label{T1IdealFluid}
\ee 
since $T_0=T(R_\star)$. In particular, evaluating at $r=R_*$, one can compute the ratio of the temperatures at $r=R_\star$ and $r=R_*$ as  
\beal{
\frac{T(R_*)}{T(R_\star)}= 1 +  G(\frac1{R_*}-\frac 1{R_\star}) \left(M_*+ \frac{6 \pi  P_0 R_*R_\star (R_*+R_\star)+2 \pi  \rho_0 \left(R_*^2 R_\star+R_*R_\star^2-2 R_*^3\right)}{3 c^2}\right).
\label{T*Tstar}
}
Note that the solutions \eqref{nu*} and \eqref{T1IdealFluid} are applicable for a case with $\rho_0, P_0$ general constants. 
 
For most observed stars, the thermal kinetic energy $k_{\text B}T$ is much smaller than the rest energy of the constituent particle $ \bar mc^2$ near surface. For instance, the temperature of the Sun at the core is estimated as of order $10^7{\rm K}\approx 1{\rm KeV}/k_{\text B}$, which is even smaller than the order of the electron mass.   
This means that if such a star would consist mainly of the ideal gas of non-relativistic particles, then it would collapse and not exist stably. Thus an ideal gas of non-relativistic particles will not be appropriate as main constituent material of stable stars.

\subsection{Analytic multilayer stellar model} 
\label{PhotonBaryon}

To the end of the construction of a model of a star with multilayer structure, finally consider the mixture of two ideal gases consisting of baryonic particles and non-relativistic ones as an example.
The ideal gas of non-relativistic particles here describes atmosphere consisting of neutral particles and ionized ones, while that of baryonic particles does main stellar constituent material. 
The total pressure and the total energy density are given by 
\beal{
\rho=\rho_{\rm b}+\rho_{\text g},~~~P=P_{\rm b}+P_{\text g},
} 
where the subscript $b$ is used for the baryonic component. 
The equation of state for an ideal gas of baryons is as usual given by  
\be 
P_{\rm b} =\hat n_{\rm b} k_{\text B} T, 
\label{IdealGasB}
\ee
while the baryonic particle number conservation implies
\be 
\hat n_{\rm b}' = \hat n_{\rm b}\frac {\rho'}{\rho + P}, 
\label{EOSBaryon}
\ee
as explained in section \ref{IF}.
Substituting \eqref{IdealGasB} into \eqref{EOSBaryon} so as to remove $\hat n_{\rm b}$ and employing \eqref{Tgradient0}, one finds $P_{\rm b}'(P_{\text g}+\rho) =P_{\rm b}(P_{\text g}+\rho)' $.
This can be solved as $P_{\rm b} = w_{\rm b}(P_{\text g}+\rho) $, where $w_{\rm b}$ is an integration constant. 
$\rho_{\rm b}$ can be determined in terms of $T$ by solving the temperature gradient equation \eqref{Tgradient0}.

For a realistic situation to a typical star, the local temperature away from the core is much smaller than the rest energy of the non-relativistic particle per the Boltzmann factor, $T\ll T_{\text g}$, where $T_{\text g}:=\frac{\bar mc^2 }{k_{\text B}}$. 
In this situation, $\rho_{\text g} \approx \bar mc^2 \hat n_{\text g} $, where $\hat n_{\text g}$ is given by \eqref{ng}, and $P_{\text g} \ll \rho_{\text g}$, so that $
P =(1+ w_{\rm b}) P_{\text g} + w_{\rm b}\rho  \approx  w_{\rm b} \rho$. This implies that in the realistic situation baryonic particles are dominant and the system can be approximately regarded as a simple fluid near and below the surface, so that the parameter $w_{\rm b}$ is related to the heat capacity ratio of the stellar material $\gamma_{\rm b}$ as $w_{\rm b} = \gamma_{\rm b} -1$, as investigated in section \ref{IF}.%
\footnote{
This consequence is indeed preferable as a property of luminous stars with taking into account that there exists considerably large convection zone underneath the solar surface \cite{RevModPhys.60.297,NASASolarPhysics}. 
}
The total energy density was determined to satisfy \eqref{Tgradient0} as $\rho \approx a_{\rm b} T^{\frac{\gamma_{\rm b}}{\gamma_{\rm b}-1}}$, where $a_{\rm b}$ is an integration constant.
To fix it, introduce the boundary of a layer of the stellar material whose radius is denoted by $R_*$ and impose the energy density of main stellar constituent material to vanish at the boundary of the layer.
\be 
\rho_{\rm b}(R_{*}) = 0.
\label{StellarSurface}
\ee
This fixes the integration constant as $a_{\rm b} \approx \rho(R_{*})/T(R_{*})^{\frac{\gamma_{\rm b}}{\gamma_{\rm b}-1}}$, where $\rho(R_{*})=\rho_{\text g}(R_{*})\approx\bar mc^2 \hat n_{\text g}(R_{*})$ and $T(R_{*})$ is the temperature at the surface of the lower layer related to the one at the surface of the upper layer as \eqref{T*Tstar}. On the other hand, the local temperature inside the star is approximated as the one to obey the power law as \eqref{TIdealFluid}, $T\approx T(R_{*})(\frac{R_{*}}{r})^{2(1 - \frac{1}{\gamma_{\rm b}}) } $, with $
T(R_{*})= \left(\frac{c^2 (\gamma_{\rm b}-1)}{2 \pi  a_{\rm b} G \left(\gamma_{\rm b}^2+4\gamma_{\rm b} -4\right)}\right)^{1 - \frac{1}{\gamma_{\rm b}}} \frac1{R_{*}^{2(1 - \frac{1}{\gamma_{\rm b}}) } }.$
Substituting the expression of $a_{\rm b}$ into this, the radius of the baryonic layer is determined as 
\be 
R_{*}\approx\sqrt{ \frac{c^2(\gamma_{\rm b}-1)}{2 \pi \rho(R_{*}) G \left(\gamma_{\rm b}^2+4\gamma_{\rm b} -4\right)} }.
\label{R*}
\ee
The physical implication of this result is twofold. 
One is that a layer with bigger density goes deeper inside the star, which is physically preferable and indeed observed as the onion structure of a star.
The other is that from this expression the deviation of the heat capacity ratio from one can be computed as 
\be 
\gamma_{\rm b} \approx 1 + 2 \pi \frac{\rho(R_{*})}{c^2} G R_{*}^2,
\label{HCR*}
\ee
where $\gamma_{\rm b}\sim1$ was used.%
\footnote{ 
For the case of the Sun, it is evaluated as $\gamma_{\rm b} -1\approx 1\times 10^{-6}$, where the following data were used: $\rho(R_*)/c^2 \to m_{\rm u}/(\frac43\pi a_{\text B}^3)\approx 2.7\times 10^3 {\rm kg/m^3}
, R_*\to \frac{R_\star}2 =\frac{R_\odot}2 = 3.5\times10^{8}\rm m$, with $a_{\text B}$ the Bohr radius, $a_{\text B}=5.3\times10^{-11}\rm m$.
This rough estimation implies that the deviation of the heat capacity ratio from one is very small. 
}
The total energy density and the baryonic component thereof are determined as 
\beal{
\rho 
\approx \rho(R_{*}) \(\frac{R_{*}}{r}\)^{2}, ~~ 
\rho_{\rm b} 
\approx\rho\times\( 1 - (\frac{R_{*}}{r})^{1-\frac3{\gamma_{\rm b}}} e^{T_{\text g}(\frac{1}{T(R_{*})}-\frac{1}{T}) }  \) . 
\label{rhorhob}
}
Then the baryonic component of pressure is given by 
$P_{\rm b} \approx w_{\rm b}\rho$, and the baryonic number density is $\hat n_{\rm b}= P_{\rm b}/k_{\text B}T$. 

These results of thermodynamic observables are valid in a parametric region with $r_{\rm H} \ll r \leq R_*, T \ll T_{\text g}$. Indeed, these expressions of thermodynamic variables develop singularity at the center of the system. This implies the existence of a core layer around the center in which thermodynamic behavior changes from the power law.
Such a core deep inside a star is expected to consist of degenerate matter \cite{10.1093/mnras/87.2.114}, so it may be very roughly approximated as a hydrostatic equilibrium state with uniform energy density. 
The pressure and temperature can be analytically computed as follows \cite{glendenning2012compact,Yokoyama:2023nld}
\beal{
P=&\frac{ \sqrt{r_{\rm c}^2-r^2}-\sqrt{r_{\rm c}^2-R_{\rm c}^2} }{3\sqrt{r_{\rm c}^2-R_{\rm c}^2} - \sqrt{r_{\rm c}^2-r^2} } \rho_{\rm c},~~
T= \frac{ 2 \sqrt{r_{\rm c}^2-R_{\rm c}^2} }{3\sqrt{r_{\rm c}^2-R_{\rm c}^2} -  \sqrt{r_{\rm c}^2-r^2} }T(R_{\rm c}), 
\label{core}
}
where $\rho_{\rm c}$ is the constant energy density, $R_{\rm c}$ is the radius of the surface of the core, $r_{\rm c} =\sqrt \frac{3}{ 8\pi G \rho_{\rm c} }$. 

Beyond the layer of main stellar material, $r>R_*$, the existing stellar matter is only the ideal gas of non-relativistic particles. Therefore, thermodynamic observables in this region are already computed in the previous section \ref{NRParticle}.

Once energy density is determined, $M_*$ and $M_\star$ are computed from \eqref{rho} as 
\be 
M_*= \frac1{c^2} \int_0^{R_*}dr 4\pi r^2 \rho, ~~ 
M_\star=M_*+ \frac1{c^2} \int_{R_*}^{R_\star}dr 4\pi r^2 \rho.
\label{M*}
\ee
The difference of these masses can be estimated as \eqref{MstarM*}. 
If the core layer is negligibly small and the energy density can be always approximated by \eqref{rhorhob} inside the star,
then $M_*$ is evaluated as
\be 
M_*\approx \frac{\rho(R_{*})}{c^2}4\pi R_{*}^3 
\label{M*2}
\ee
This has a simple physical meaning that the total mass of a star can be estimated just by the product of mass density at the surface of the baryonic layer and its volume up to a numerical factor $3$. 
To confirm validity of this expression, below estimate the solar mass by using it.   
To this end, assume that the contribution of the atmosphere to the mass is negligible, $M_\star \approx M_*$, and that the radius of the upper layer is set to be identical to the solar one related to that of the lower layer as $R_\star\equiv R_\odot = \eta R_*$, where $\eta$ is a numerical factor greater than 1.
On the other hand, the mass density at the surface of the baryonic layer is estimated as $\rho(R_{*})/c^2 = \bar m \hat n_{\text g}(R_*) = \mu_{\rm e} m_{\rm u}/(\frac43 \pi a_{\text B})^3 $, where the number density is $\hat n_{\text g}(R_*) \approx 1/(\frac43 \pi a_{\text B})^3 $ with $a_{\text B}$ the Bohr radius and $\mu_{\rm e}\approx1.2$ since the weight fraction of hydrogen is $X=0.6$ for the case of the Sun \cite{pols2011stellar}.\footnote{This approximation means that the Sun is treated almost as a dense pack of hydrogens at surface.} From these data, the solar gravitational mass is evaluated as 
\be 
M_\star\approx \frac{\mu_{\rm e} m_{\rm u}}{\frac43 \pi a_{\text B}^3} \times 4\pi  (\frac{R_\odot}{\eta})^3 
\approx \frac{1}{\eta^3}\times 10^{34}\rm g. 
\label{SolarMass}
\ee
This estimation yields the same order of the solar mass, $ M_\odot= 2.0\times 10^{33}\rm g$. 

In order to construct a model of a luminous star, consider energy transported by radiation measured by the luminosity $L_r$.
In this paper, the luminosity is literally defined as the energy of radiation only flowing outwards across the sphere of radius $r$ per unit time here.%
\footnote{ 
In some contexts, the luminosity is defined as the net energy flow transported not only by radiation but also by convection in a more involved context such as the mixing length theory. The reason to adopt the definition in this paper is because the net energy flow will be evaluated by using entropy as $TdS$. 
}
This is evaluated in the background metric \eqref{metric} as 
\be 
L_r= 4\pi r^2 e^{\lambda/2} F, 
\label{LF}
\ee 
where $F$ is the radiative energy flux summed all over the frequencies $F=\int d\nu F_\nu$ and $F_\nu$ is the energy flux component of radiation with frequency $\nu$.
Now assume that the source of radiation and the absorption body are absent.
Then the energy flux component $F_\nu$ is related to its pressure $P_\nu$ by $
{dP_\nu\over d\tau_\nu}= \frac {F_\nu}c,$
where $\tau_\nu$ is the frequency-sensitive optical depth defined through the absorption coefficient $\alpha_\nu$ as $d\tau_\nu = - \alpha_\nu dr. $
Therefore $ {dP_\nu\over dr}=-\alpha_\nu \frac {F_\nu}c$. 
Summing all over the modes labeled by the frequency $\nu$ leads to 
\be 
P_{\rm rad}'=-\alpha_{\rm R} \frac {F}c, 
\label{PRadiation'}
\ee
where $\alpha_{\rm R}$ is an averaged absorption coefficient called the Rosseland mean.
The radiative component of pressure is given by $P_{\rm rad}=\frac{4\sigma}{3c}T^4$, so $T'=\frac{3c}{16\sigma T^3} P_{\rm rad}'$. 
Substituting \eqref{PRadiation'} into this yields 
\be 
T'=-\frac{3\alpha_{\rm R}}{16\sigma T^3} F.
\label{T'F}
\ee
Then using \eqref{LF} one finds the relation between the temperature gradient and the local luminosity as 
\beal{
T'
=-\frac{3\alpha_{\rm R}}{16\sigma T^3 } \frac{L_r}{4\pi r^2(1 - {2 G\check E_r \over rc^2})^{-\half} }. 
\label{T'LGR}
} 
This is the relativistic extension of the temperature gradient equation \eqref{T'Radiation} for the radiation dominant zone.
As a byproduct, one can derive a relativistic result on the luminosity by plugging \eqref{Tgradient} into \eqref{T'LGR} as 
\beal{  
L_r 
= \frac{64\pi\sigma  }{3c^2\alpha_{\rm R}} \frac{ G( \check E_r+4\pi r^3 P)}{ (1-2\frac{ G\check E_r }{rc^2})^{3/2} } T^4. 
\label{luminosityGR}
}
Indeed, near the stellar surface, \eqref{T'LGR} reduces to \eqref{T'Radiation} with taking into account the relation between the Rosseland mean and the opacity as $\alpha_{\rm R} \approx \varrho \kappa $ with $\varrho$ the mass density of stellar matter, as asserted before, while \eqref{luminosityGR} reduces to 
\beal{
L_r 
\approx \frac{64\pi\sigma  }{3c^2\alpha_{\rm R}} G \check E_r T^4. 
\label{luminosityNR}
}
Note that these expressions for the luminosity are valid in the region with matter present. In the matter empty region, the Rosseland mean and opacity vanish, and from \eqref{T'F}, the temperature is constant. This is consistent with the result of the local temperature in the matter empty region \eqref{Tstar} effectively.

To extract physical implication, evaluate \eqref{luminosityNR} in the atmospheric layer, $R_* \leq r \leq R_{\star}$.
Near the upper surface, $r\lesssim R_\star$, the pressure is negligible as long as \eqref{radiusCondition} is satisfied, and $\check E_r$ and $T$ are almost constant from \eqref{Er} and \eqref{Tstar}, respectively, as seen in section \ref{StellarTemperature}. 
Thus 
\be 
L_r
\approx \frac{64\pi\sigma }{3\alpha_{\rm R\star}} G M_\star T(R_\star)^4.
\ee
Here $\alpha_{R\star}$ is the Rosseland mean near the upper surface and becomes a constant. 
This shows that the luminosity in this region is also constant, which is consistent with the Stefan-Boltzmann law, $L_r \approx L_\star$, where 
\be 
L_\star = 4\pi R_{\star}^2 \sigma T_\star^4 . 
\label{SB}
\ee 
Here the radius $R_{\star}$, which was introduced so as to satisfy the condition \eqref{radiusCondition}, is defined for the Stefan-Boltzmann law to hold at the radius, and $T_\star$ is the effective surface temperature identified with $T_\star=T(R_\star)$. Then a novel relation is obtained as 
\be 
R_{\star}^2=\frac{16 G M_\star}{3\alpha_{\rm R\star}}.
\label{Rstar}
\ee
This can be used to estimate the Rosseland mean at the measured point.\footnote{ 
The expression \eqref{Rstar} cannot be used in a region of stellar corona, since, as argued in section \ref{StellarCorona}, the Newtonian approximation is not valid any more in a plasma state of hydrostatic equilibrium. In order to estimate the Rosseland mean in stellar corona, one has to use an expression obtained without using the Newtonian approximation such that $\alpha_{\rm R\star}=\frac{16 G (M_\star+4\pi R_{\star}^3P(R_\star)/c^2)}{3R_{\star}^2}$.
}
It would be interesting to compare the estimated result and detailed data of the opacity \cite{1957CoKon..42..109K,shapiro2008black,2013sse..book.....K}. 

\subsubsection{Analytic multilayer solar model} 
\label{SolarModel} 

In order to build a concrete model of a star from the above multilayer stellar model, internal data of the star is necessary.  
To illustrate how to build such a concrete model, here construct a multilayer model of the Sun as an example. It is known that the Sun has a convection zone up to around three quarters of the total radius from the surface and the deeper rest is a radiation zone except the core \cite{RevModPhys.60.297,NASASolarPhysics}. 
Here, for simplicity, the region of solar corona is neglected for simplicity, and the convective zone and the radiative one are both realized by a baryonic layer consisting of an ideal gas of baryonic particles with the heat capacity ratio varying in the layer. 
Since the total solar mass is already estimated correctly as seen in \eqref{SolarMass}, below investigate behaviors of local observables. 

The surface of the atmosphere starts from the radius $R_\star$, which is set to the solar one, $R_\star:=R_\odot$. The layer of atmosphere continues roughly up to the half of the solar radius, $R_*=\half R_\star$, from which the baryonic layer starts. The baryonic layer is roughly divided into a convective zone, which ends at the radius $\tilde R_* = \frac34 R_*$, and a radiative one, which ends at $r=R_{\rm c}$ and connects to the core smoothly. Here the radius of the core is chosen as $R_{\rm c}:=5r_{\rm H}$ by hand for illustration.
Accordingly the heat capacity ratio starts with $1$ at the surface of the baryonic layer and increases as it goes deeper. It takes the value $4/3$ at the transition point from the convective zone to the radiative zone, and $5/3$ at the one from the baryonic layer to the core, since the core is expected to be supported by repulsive force of degenerate electron gas \cite{10.1093/mnras/87.2.114}, whose heat capacity ratio is $5/3$ \cite{prialnik2000introduction}.
Note that this realization of the radiation zone is different from the traditional one, because the temperature gradient equation obeys \eqref{Convection} in an ideal gas of baryonic particles, which is supposed to hold not in a radiative zone but in a convective one.   
There are two reasons for this. One is that, as argued in section \ref{application2radiation}, the presented relativistic hydrostatic structure equations cannot be applied for null fluid. This does not allow one to solve the structure equations combining with the equation \eqref{T'LGR}. 
The other is that if only the equation \eqref{T'LGR} is used as the stellar structure equations without using \eqref{Convection}, then one of the relativistic structure equations \eqref{Tgradient0}, which is expected to always hold in local thermodynamic equilibrium, is not satisfied.
On these reasons, the radiative zone is realized by a simple fluid to satisfy the equation \eqref{Convection} with $w=1/3$ in this model. 

\begin{figure}[tbh]
  \begin{center}
  \subfigure[Solar multilayer structure.]{\includegraphics[scale=.4]{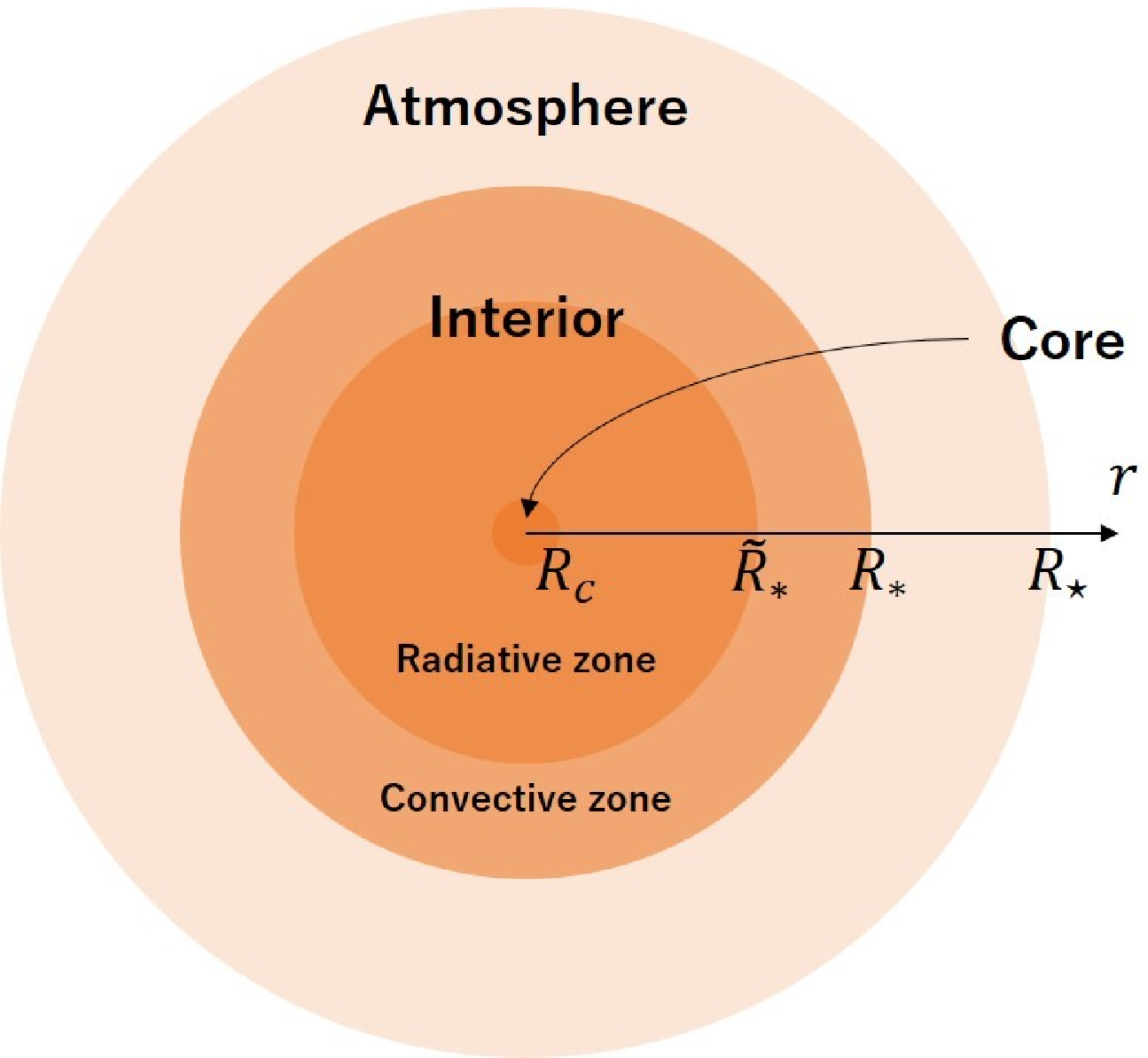}
\label{SchematicSun}
  }
  \qquad\qquad
  \subfigure[Behaviors of local observables.]{\includegraphics[scale=.6]{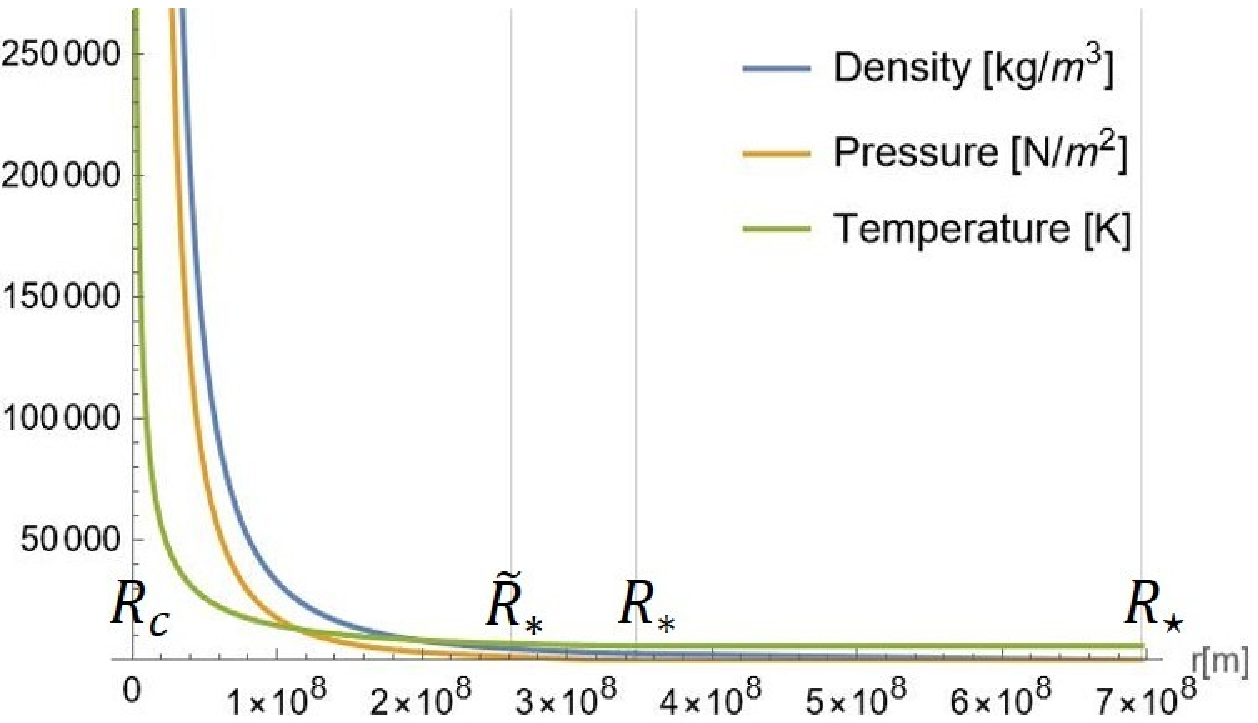}
\label{SolarObservables}
  }
  \end{center}
  \vspace{-0.5cm}
  \caption{(a) A schematic picture of solar multilayer structure and (b) the behaviors of local thermodynamic observables are drawn. Three local observables are plotted in the same graph even though they have different units in order to describe their behaviors concisely.}
  \label{Sun}
\end{figure}

The behaviors of local observables inside the Sun are summarized in table~\ref{StellarModel} and figure~\ref{Sun}.
Their characteristic values are easily calculated as follows.  
On the local temperature, as an initial condition, $T_\star$ is set to the solar surface temperature observed by an asymptotic observer in accord with the Stefan-Boltzmann law, $T_\star:=T_\odot \approx 6\times 10^3 {\rm K}$. The temperature does not change so much in the layer of atmosphere, so that it lasts at the surface of the baryonic layer, $T(R_*) \approx T_\odot$. Then it increases as it goes deeper. 
At the transition point from the convective zone to the radiative one, $T(\tilde R_{*})=T(R_*)\left(\frac{R_{*}}{\tilde R_{*}}\right)^{1/2} \approx 7\times 10^3 {\rm K}$, while at the point from the baryonic layer to the core $T(R_{\rm c})=T(R_{*})\left(\frac{R_{*}}{R_{\rm c}}\right)^{5/4} \approx 2\times10^7{\rm K}$. Note that in the current case, $r_{\rm c}\gg R_{\rm c}$, so the temperature of the center of the system is almost equal to the one at the surface of the core, $T(0)\approx T(R_{\rm c})$. The core temperature estimated here is of the same order as the earlier one obtained by using the classic stellar structure equation \cite{RevModPhys.60.297,carroll2007introduction}.

On the density, an initial condition is given at the boundary between the atmosphere and the interior, $r=R_*$, where the density is estimated as a dense pack of hydrogens, $\rho(R_*)/c^2\approx m_{\rm u}/(\frac43\pi a_{\text B}^3)\approx 3\times 10^3 {\rm kg/m^3}$.
Then at the transition point from the convective zone to the radiative one $\rho(\tilde R_{*})/c^2=\rho(R_{*})/c^2\(\frac{R_*}{\tilde R_{*}}\right)^2 \approx 5\times 10^{3} {\rm kg/m^3}$, while at the core $\rho_{\rm c}/c^2=\rho(R_{*})/c^2\(\frac{R_*}{R_{\rm c}}\right)^2 \approx 1\times 10^{12} {\rm kg/m^3}$. This density at the core is much bigger than the earlier result around $1.5 \times 10^{5} {\rm kg/m^3}$ 
\cite{RevModPhys.60.297,carroll2007introduction}. In fact, the value is still bigger than the known average density of a typical white dwarf around $3\times10^9 {\rm kg/m^3}$ but less than that of a neutron star around $5\times10^{14} {\rm kg/m^3}$ \cite{glendenning2012compact}.
However, this result of high density is in some sense reasonable because the core is assumed to be supported by repulsive force of degenerate electrons so its density  is expected to be comparable to that of a white dwarf. 
Accordingly the pressure in the model also becomes much higher than the earlier result. 
The behaviors of local observables for this solar model are plotted in figure \ref{Sun} (b) by determining the heat capacity ratio in each zone by the linear interpolation using the boundary values.
They are qualitatively different from the earlier ones \cite{RevModPhys.60.297,carroll2007introduction}.
This qualitative difference presumably arises due to that of the realization of the radiative zone. 

It is important to stress that the values obtained here are very rough without fine tuning. Therefore they can be possibly made more precise by tuning the parameter with more precise data. 
Such a precise evaluation and further discussion of physical interpretations of these differences are left to future work.

\subsubsection{Comment on the Eddington bound} 
\label{EddingtonLimit} 

In the above model of luminous stars, the radiation pressure is assumed to be too small to involve hydrostatic structure \cite{1957CoKon..42..109K}.   
This approximated treatment is not used in the mixing length theory \cite{1948ZA.....25..135B,vitense1953wasserstoffkonvektionszone}, in which both the convective energy flux and the radiant energy flux  involve the traditional hydrostatic structure equations. (See also \cite{osti_4807566}.)
This approximation will be valid as long as the outward force caused by radiation is sufficiently small compared to the gravitational one produced by matter inside the star: 
\be 
|{dP_{\rm rad} \over dr}| \ll |{dP\over dr}|. 
\ee
Using \eqref{LF} and \eqref{luminosityGR}, one can compute the left-hand side as $|P_{\rm rad}'|= \frac { \alpha_{\rm R} L_r}{4\pi r^2 (1-2\frac{ G\check E_r }{rc^2})^{-\half}c}$, while the right-hand is already computed as \eqref{TOV}. 
Therefore the above inequality becomes 
\be 
L_r \ll (1+w_{\rm b})\rho \frac{4\pi G( \check E_r+4\pi r^3 P)}{\alpha_{\rm R} c (1-2\frac{ G\check E_r }{rc^2})^{\frac32} },
\ee 
where $P \approx w_{\rm b} \rho$ was used. 
Further estimation of the right-hand side needs to specify the expression of the Rosseland mean. 
In a non-relativistic hydrostatic system, the Rosseland mean takes the form of the product of the mass density $\varrho$ and the opacity $\kappa$. In a general relativistic hydrostatic system, the energy density $\rho$ is more fundamental than the mass density $\varrho$, so the Rosseland mean is defined by using the energy density as $\alpha_{\rm R}= \kappa \rho/c^2$.
Then the above can be rewritten as 
\be 
L_r \ll \gamma_{\rm b} \frac{4\pi cG( \check E_r+4\pi r^3 P)}{\kappa (1-2\frac{ G\check E_r }{rc^2})^{\frac32} }.
\label{EddingtonGR}
\ee 
The right-hand side corresponds to the relativistic extension of the Eddington limit for this stellar model. 
Indeed, evaluating this around a stellar surface with the Newtonian approximation valid, one finds
\be 
L_{R_\star} \ll \frac{4\pi cG_{\rm N} M_\star}{\kappa}. 
\label{EddingtonNR}
\ee 

\section{Discussion}
\label{Discussion}

The relativistic extension of the classic stellar structure equations has been investigated in the Einstein gravity.
It has been pointed out that the TOV equation and two gradient ones for local gravitational mass and the Tolman temperature form a closed set of differential equations, and the set has been put forward as the desired relativistic extension of the stellar structure equations. The proposed relativistic hydrostatic structure equations have been shown to be endorsed by laws of local thermodynamics.%
\footnote{ 
This point should not be overlooked for a precise model building of a celestial body. A certain polytropic relation is often assumed to study a stellar structure in classic papers such as \cite{1935MNRAS..95..207C} and standard textbooks on astrophysics, but it is not clear or described whether it is indeed endorsed by laws of local thermodynamics. If it is not, then such a model is a rough one which awaits replacement by a thermodynamically correct one, even though it has contributed enormously to the understanding of physics of an astronomical body. 
}
From them, the exact form of the relativistic Poisson equation has been derived, and it has been converted into the steady-state heat conduction equation holding non-perturbatively with respect to the Newton constant by using the relation between the gravitational potential and the Tolman temperature. 

A couple of applications have been presented. 
One is to a hydrostatic equilibrium system consisting of a single ideal gas of particles with the particle number current conserved called baryonic particles. In this system, the proposed temperature gradient equation reduces precisely to the conventional one known in the convection zone. It has been predicted that the heat capacity ratio almost becomes one in the Newtonian convective regime such as the neighborhood of the solar surface.  
The generalized steady-state heat conduction equation has been solved exactly and thermodynamic observables determined non-perturbatively in the Newton constant, which exhibit the power law behavior such that $T\propto r^{-2(\gamma-1)\over \gamma}, \rho\propto p\propto r^{-2}$, with $\gamma$ the heat capacity ratio. This result implies that this model is applicable to a plasma state and thus to stellar corona.  
Another application is to a hydrostatic equilibrium system of an ideal gas of non-relativistic particles, for which the local temperature has been determined perturbatively in the Newton constant. 
Finally, by combining two ideal gases of baryonic particles and non-relativistic ones, an analytic multilayer structure of luminous stars has been revealed with a simple example. 
In this model, the interior consists mainly of the ideal gas of baryonic particles, while the atmosphere is constituted by non-relativistic particles. 
This model also admits the layer of degenerate core and the stellar corona by changing the constituents suitably. 
By coupling the system to radiation, the traditional temperature gradient equation in a radiation zone has been extended to its relativistic version. 
Inputing observational data into the model, a concrete model of the Sun has been constructed and analytic expressions of thermodynamic observables have been determined.
Finally, the Eddington bound has been obtained as a condition to justify the approximation such that the contribution of radiation is too small to be involved in the structure equations. 

In the proposed hydrostatic structure equations, 
the relativistic extension for the luminosity gradient \eqref{Lr'} is not included.
This is simply because structure equations are closed without the variable of luminosity. This implies that the variables on rate are not fundamental for relativistic hydrostatic equilibrium. This conclusion is actually preferable with taking into account the existence of a non-radiating star which can be described by a relativistic hydrostatic equilibrium system. 
However, this does not mean that the gradient equation for luminosity is not useful in a relativistic hydrostatic equilibrium system. 
Its naive relativistic extension will be given by 
\beal{
{d L_r\over d r}=& \frac{4\pi r^2}{ (1-2\frac{ G\check E_r }{rc^2})^{\frac12} } \rho \epsilon, 
\label{Lr'GR}
}
which could be useful to extract information of the radiant energy production rate $\epsilon$ per unit energy for a hydrostatic stellar system with radiation. 
The information of energy production rate becomes important to build a model in detail \cite{RevModPhys.60.297}, particularly to investigate an evolutionary process of a star as a non-equilibrium open system including dissipation. Further extension including such a hydrodynamical process is an interesting future work. 

If the relativistic hydrostatic structure equations with a simple fluid is applied to stellar corona, then pressure vanishes at the far asymptotic region.
This result conflicts with the earlier one obtained by using the non-relativistic stellar structure equations, and what is a flaw in the earlier argument is ignoring a term containing general relativistic effect in the stellar equilibrium equation which cannot be neglected in a situation with pressure finite in the far asymptotic region. It would be interesting to investigate this model further to solve problems known in solar corona  \cite{RevModPhys.60.297,NASASolarPhysics}.  

In the multilayer stellar model presented in this paper, it consists of two ideals gases of baryonic particles and non-relativistic ones as a simple example. It is also possible to pile up a similar layer or another one consisting of a different type of fluid. Such a multilayer structure should be arranged to build a model for each observed star. 
In the analysis, radiation does not involve in the structure equations but just plays a role of a probe of local temperature. 
This approximated treatment of radiation is inevitable since the proposed hydrostatic structure equations, and thus the classic stellar ones, cannot be applied to null fluid, strictly speaking, and it is justified as long as the Eddington bound is sufficiently satisfied. However the radiative contribution may not be negligible inside the star, in particular, with so high temperature that an effective method of analysis is only numerical calculation. In fact, it is expected that there exists a large radiative zone in the interior of the Sun \cite{RevModPhys.60.297,NASASolarPhysics}. In this paper, the radiative zone has been realized by a simple fluid with $w=1/3$ as a provisional treatment. 
It is important to extend the presented relativistic hydrostatic structure equations so as to include the radiative contribution and test the validity of the approximation used in this paper by estimating the radiative contribution. 

The application of the proposed relativistic hydrostatic structure equations has been focused on the construction of a model of luminous stars. It would be tempted to apply them to a different type of stars such as a compact star consisting of degenerate matter and to investigate the finite temperature effect. A caveat for the application of the relativistic hydrostatic structure equations to such compact stars is to include the local chemical potential and temperature so as to be consistent with the local thermodynamic relations. It would be of a great interest to investigate how the thermodynamic relations shown in \cite{Yokoyama:2023nld} including chemical potential in such a system is derived and whether such thermal interaction gives rise to any significant effect on the system.

The author hopes to address these issues and will report the progress in the near future. 

\section*{Acknowledgement}
The author would like to thank Masaki Mori for a helpful comment on the first version. 
This work is supported in part by the Grant-in-Aid of the Japanese Ministry of Education, Sciences and Technology, Sports and Culture (MEXT) for Scientific Research (No.~JP22K03596). 

\bibliographystyle{utphys}
\bibliography{stst}

\providecommand{\href}[2]{#2}\begingroup\raggedright\begin{thebibliography}{10}

\bibitem{HelmholtzLXIVOT}
H.~L.~F. von Helmholtz, ``Lxiv. on the interaction of natural forces,'' {\em
  Philosophical Magazine Series 1} {\bf 11} 489--518.

\bibitem{kelvin1889popular}
W.~Kelvin, {\em Popular Lectures and Addresses}.
\newblock Nature series. Macmillan and Company, 1889.

\bibitem{Bahcall:2000xc}
J.~N. Bahcall, ``{How the sun shines},'' {\em J. Roy. Astron. Soc. Canada} {\bf
  94} (2000) 219, \href{http://arXiv.org/abs/astro-ph/0009259}{{\tt
  astro-ph/0009259}}.

\bibitem{eddingtoninternal}
A.~Eddington, {\em The Internal Constitution of the Stars}.
\newblock Cambridge University Press/Dover Publications/Cambridge University
  Press, 1930/1959/1988.

\bibitem{PhysRev.53.595}
G.~Gamow, ``Nuclear energy sources and stellar evolution,'' {\em Phys. Rev.}
  {\bf 53} (Apr, 1938) 595--604.

\bibitem{Bethe:1939bt}
H.~A. Bethe, ``{Energy production in stars},'' {\em Phys. Rev.} {\bf 55} (1939)
  434--456.

\bibitem{harwit1988astrophysical}
M.~Harwit, {\em Astrophysical Concepts}.
\newblock Astronomy and Astrophysics Library. Springer New York, 1988.

\bibitem{phillips1994physics}
A.~Phillips, {\em The Physics of Stars}.
\newblock Manchester Physics Series. Wiley, 1994.

\bibitem{tayler1994stars}
R.~Tayler, {\em The Stars: Their Structure and Evolution}.
\newblock Cambridge University Press, 1994.

\bibitem{rnaasaaaa67bib2}
C.~J. Hansen and S.~D. Kawaler, {\em Stellar Interiors}.
\newblock Springer, 1994.

\bibitem{2013sse..book.....K}
R.~{Kippenhahn}, A.~{Weigert}, and A.~{Weiss}, {\em {Stellar Structure and
  Evolution}}.
\newblock 2013.

\bibitem{choudhuri2010astrophysics}
A.~Choudhuri, {\em Astrophysics for Physicists}.
\newblock Astrophysics for Physicists. Cambridge University Press, 2010.

\bibitem{1938ApJ....88..472K}
G.~P. {Kuiper}, ``{The Empirical Mass-Luminosity Relation.},'' {\em
  Astrophysical Journal} {\bf 88} (Nov., 1938) 472.

\bibitem{rnaasaaaa67bib5}
A.~W. Mann, E.~Gaidos, and M.~Ansdell, ``Spectro-thermometry of m dwarfs and
  their candidate planets: Too hot, too cool, or just right?,'' {\em ApJ} {\bf
  779} (2013) 188.

\bibitem{Cuntz_2018}
M.~Cuntz and Z.~Wang, ``The mass–luminosity relation for a refined set of
  late-k/m dwarfs,'' {\em Research Notes of the AAS} {\bf 2} (jan, 2018) 19.

\bibitem{Misner:1974qy}
C.~W. Misner, K.~S. Thorne, and J.~A. Wheeler, {\em {Gravitation}}.
\newblock W. H. Freeman, San Francisco, 1973.

\bibitem{PhysRev.55.374}
J.~R. Oppenheimer and G.~M. Volkoff, ``On massive neutron cores,'' {\em Phys.
  Rev.} {\bf 55} (Feb, 1939) 374--381.

\bibitem{osti_4092046}
G.~S. Saakyan and M.~A. Mnatsakanyan, ``Neutron configurations in the
  generalized newtonian theory of gravitation.,'' {\em Astrophysics (Engl.
  Transl.) 4: 62-7(Summer 1968).}

\bibitem{1964SvA.....8..147S}
G.~S. {Saakyan} and Y.~L. {Vartanyan}, ``{Basic Parameters of Baryon
  Configurations},'' {\em Soviet Astronomy} {\bf 8} (Oct., 1964) 147.

\bibitem{hartle2003gravity}
J.~Hartle, {\em Gravity: An Introduction to Einstein's General Relativity}.
\newblock Addison-Wesley, 2003.

\bibitem{10.1093/mnras/87.2.114}
R.~H. Fowler, ``{On Dense Matter},'' {\em Monthly Notices of the Royal
  Astronomical Society} {\bf 87} (12, 1926) 114--122,
  \href{http://arXiv.org/abs/https://academic.oup.com/mnras/article-pdf/87/2/114/3623303/mnras87-0114.pdf}{{\tt
  https://academic.oup.com/mnras/article-pdf/87/2/114/3623303/mnras87-0114.pdf}}.

\bibitem{zel2014stars}
Y.~Zel'dovich and I.~Novikov, {\em Stars and Relativity}.
\newblock Dover Books on Physics. Dover Publications, 2014.

\bibitem{Yokoyama:2023nld}
S.~Yokoyama, ``{Local Thermodynamics and Entropy for Relativistic Hydrostatic
  Equilibrium},'' \href{http://arXiv.org/abs/2304.06196}{{\tt 2304.06196}}.

\bibitem{Aoki:2020nzm}
S.~Aoki, T.~Onogi, and S.~Yokoyama, ``{Charge conservation, entropy current and
  gravitation},'' {\em Int. J. Mod. Phys. A} {\bf 36} (2021), no.~29, 2150201,
  \href{http://arXiv.org/abs/2010.07660}{{\tt 2010.07660}}.

\bibitem{PhysRev.35.904}
R.~C. Tolman, ``On the weight of heat and thermal equilibrium in general
  relativity,'' {\em Phys. Rev.} {\bf 35} (Apr, 1930) 904--924.

\bibitem{1948ZA.....25..135B}
L.~{Biermann}, ``{Konvektion in rotierenden Sternen},'' {\em Z. Astrophys} {\bf
  25} (Jan., 1948) 135.

\bibitem{vitense1953wasserstoffkonvektionszone}
E.~Vitense, ``Die wasserstoffkonvektionszone der sonne. mit 11
  textabbildungen,'' {\em Z. Astrophys} {\bf 32} (1953) 135.

\bibitem{1958ApJ...128..664P}
E.~N. {Parker}, ``{Dynamics of the Interplanetary Gas and Magnetic Fields.},''
  {\em Astrophysical Journal} {\bf 128} (Nov., 1958) 664.

\bibitem{1935ApJ....82..435H}
O.~{Heckmann}, ``{REVIEW: Relativity, Thermodynamics and Cosmology, by R. C.
  Tolman},'' {\em apj} {\bf 82} (Dec., 1935) 435.

\bibitem{Hawking:1973uf}
S.~W. Hawking and G.~F.~R. Ellis, {\em {The Large Scale Structure of
  Space-Time}}.
\newblock Cambridge Monographs on Mathematical Physics. Cambridge University
  Press, 2, 2011.

\bibitem{Aoki:2020prb}
S.~Aoki, T.~Onogi, and S.~Yokoyama, ``{Conserved charges in general
  relativity},'' {\em Int. J. Mod. Phys. A} {\bf 36} (2021), no.~10, 2150098,
  \href{http://arXiv.org/abs/2005.13233}{{\tt 2005.13233}}.

\bibitem{1935MNRAS..95..207C}
S.~{Chandrasekhar}, ``{The highly collapsed configurations of a stellar mass
  (Second paper)},'' {\em Monthly Notices of the Royal Astronomical Society,
  Vol. 95, p.207-225} {\bf 95} (Jan., 1935) 207--225.

\bibitem{1906Schwarzschild}
K.~{Schwarzschild}, ``{On the equilibrium of the Sun's atmosphere},'' {\em
  Nachrichten von der K{\"o}niglichen Gesellschaft der Wissenschaften zu
  G{\"o}ttingen. Math.-phys. Klasse} {\bf 195} (jan, 1906) 41--53.

\bibitem{RevModPhys.60.297}
J.~N. Bahcall and R.~K. Ulrich, ``Solar models, neutrino experiments, and
  helioseismology,'' {\em Rev. Mod. Phys.} {\bf 60} (Apr, 1988) 297--372.

\bibitem{NASASolarPhysics}
NASA, ``Solar physics,''
  \href{http://arXiv.org/abs/https://solarscience.msfc.nasa.gov/interior.shtml}{{\tt
  https://solarscience.msfc.nasa.gov/interior.shtml}}.

\bibitem{1951ZA.....29..274B}
L.~{Biermann}, ``{Kometenschweife und solare Korpuskularstrahlung},'' {\em Z.
  Astrophys} {\bf 29} (Jan., 1951) 274.

\bibitem{1957Obs....77..109B}
L.~{Biermann}, ``{Solar corpuscular radiation and the interplanetary gas},''
  {\em The Observatory} {\bf 77} (June, 1957) 109--110.

\bibitem{1960SPhD....5..361G}
K.~I. {Gringauz}, V.~V. {Bezrokikh}, V.~D. {Ozerov}, and R.~E. {Rybchinskii},
  ``{A Study of the Interplanetary Ionized Gas, High-Energy Electrons and
  Corpuscular Radiation from the Sun by Means of the Three-Electrode Trap for
  Charged Particles on the Second Soviet Cosmic Rocket},'' {\em Soviet Physics
  Doklady} {\bf 5} (Sept., 1960) 361.

\bibitem{10.1007/978-94-011-7542-5_7}
C.~W. Snyder and M.~Neugebauer, ``Interplanetary solar-wind measurements by
  mariner ii,'' in {\em Proceedings of the Plasma Space Science Symposium},
  C.~C. Chang and S.~S. Huang, eds., pp.~67--90.
\newblock Springer Netherlands, Dordrecht, 1965.

\bibitem{verscharen2019multiscale}
D.~Verscharen, K.~G. Klein, and B.~A. Maruca, ``The multi-scale nature of the
  solar wind,'' 2019.

\bibitem{glendenning2012compact}
N.~Glendenning, {\em Compact Stars: Nuclear Physics, Particle Physics and
  General Relativity}.
\newblock Astronomy and Astrophysics Library. Springer New York, 2012.

\bibitem{pols2011stellar}
O.~Pols, {\em Stellar Structure and Evolution}.
\newblock Astronomical Institute Utrecht, 2011.

\bibitem{1957CoKon..42..109K}
S.-M. {Kung}, ``{The opacity and the internal structure of the Sun},'' {\em
  Commmunications of the Konkoly Observatory Hungary} {\bf 42} (Jan., 1957)
  109.

\bibitem{shapiro2008black}
S.~Shapiro and S.~Teukolsky, {\em Black Holes, White Dwarfs, and Neutron Stars:
  The Physics of Compact Objects}.
\newblock Wiley, 2008.

\bibitem{prialnik2000introduction}
D.~Prialnik, {\em An Introduction to the Theory of Stellar Structure and
  Evolution}.
\newblock Cambridge University Press, 2000.

\bibitem{carroll2007introduction}
B.~Carroll and D.~Ostlie, {\em An Introduction to Modern Astrophysics}.
\newblock Pearson Addison-Wesley, 2007.

\bibitem{osti_4807566}
J.~P. Cox, ``Principles of stellar structure. volume 2. applications to
  stars.,''.

\end{thebibliography}\endgroup

\end{document}